\documentclass{moriond}

\def\Journal#1#2#3#4{{#1} {\bf #2}, #3 (#4)}

\def\PRL{\em Phys. Rev. Lett.}

\def\be{\begin{equation}}
\def\ee{\end{equation}}
\def\bea{\begin{eqnarray}}
\def\eea{\end{eqnarray}}

\newcommand{\Photo}{}

\begin{document}
\vspace*{4cm}
\title{THE BEAUTY OF THE RARE: $B_{(s)}\rightarrow \mu^{+} \mu ^{-}$ AT THE LHCb}

\author{ S. FERRERES SOLÉ \\ 
on behalf of the LHCb collaboration }

\address{Nikhef, Science Park 105, \\
1098 XG Amsterdam, Netherlands}

\maketitle\abstracts{The latest and improved measurements in 
$B^0_{(s)}\rightarrow\mu^+ \mu^-$ processes have been performed by LHCb 
using full Run 1 and Run 2 data. The measurement of the branching fraction of 
$B^0_{s}\rightarrow\mu^+ \mu^-$ is presented as the most precise single experiment 
measurement of such a rare process. While no significant excess is found for 
$B^0\rightarrow\mu^+ \mu^-$, an stringent upper limit is set for its 
braching fraction. Both results are in agreement with SM predictions and with 
previous results.}

\section{Theoretical beauty}
Flavour changing neutral currents (FCNC) are suppressed in the SM for several reasons. 
They are forbidden at tree level and can only take place through loop transitions involving 
multiple weak interactions. This adds to their branching fraction calculation loop and 
weak interaction suppression factors. Apart from this, they are also affected by the GIM 
mechanism, arising from the unitarity property of the CKM matrix.

The $B_{s}^0\rightarrow\mu^+\mu^-$ and $B^0 \rightarrow \mu ^+ \mu ^-$ decays, 
which will be further referred to as $B_{(s)}^0\rightarrow\mu^+\mu^-$, fall under this 
category. Such decays are even further affected by helicity suppression, originating from 
the relation between the spin of the $B_{(s)}^0$ and the helicity of the muons. 

These processes can be described using effective field theory. Their decay amplitude can 
be written using an effective Hamiltonian as \cite{buras}:
\begin{equation}
       \mathcal{M}(B_{(s)}^0 \rightarrow \mu^+\mu^-) = 
       \langle \mu^+\mu^- | \mathcal{H}_{eff} | B_{(s)}^0 \rangle \propto
       \sum_{i = 10, S, P} C_{i}\mbox{ }\langle \mu^+\mu^- | \mathcal{O}_i | B_{(s)}^0 \rangle
       \label{eq:matrix_element}
\end{equation}
where $C_{i}$ are the perturbative Wilson coefficients and $\mathcal{O}_i$ are the non-perturbative 
Wilson operators, describing the short and long distance effects, respectively. 
The $B_{(s)}^0\rightarrow\mu^+\mu^-$ final state is purely leptonic whereas the initial state is 
purely hadronic, resulting in a lack of direct interaction between the two. Consequently, 
the matrix element in Eq.~\ref{eq:matrix_element} can be factorised into the hadronic and 
leptonic parts, resulting in a very clean theoretical calculation:
\begin{equation}
       \mathcal{M}(B_{(s)}^0 \rightarrow \mu^+\mu^-) = 
       \sum_{i = 10, S, P} C_{i}\mbox{ }\langle \mu^+\mu^- | \mathcal{O}_{i,ll} | 0 \rangle \otimes 
       \langle 0 | \mathcal{O}_{i, qq} | B_{(s)}^0 \rangle
\end{equation}

All these properties make $B_{(s)}^0\rightarrow\mu^+\mu^-$ decays extremely rare in 
the SM with very precise predicted time-integrated branching fraction \cite{beneke}:
\bea
\mathcal{B}(B_s^0 \rightarrow \mu^+ \mu^-) & = &(3.66 \pm 0.14)\times 10^{-9} \nonumber\\
\mathcal{B}(B^0 \rightarrow \mu^+ \mu^-) & = &(1.03 \pm 0.05)\times 10^{-10} \nonumber
\eea

As a final remark, only the Wilson operator $\mathcal{O}_{10}$ describing axial vector 
interactions contributes to the $B_{(s)}^0\rightarrow\mu^+\mu^-$ processes in the SM. 
While this operator is strongly affected by helicity suppression, the scalar and 
pseudo-scalar contributions, $\mathcal{O}_S$ and $\mathcal{O}_P$, are not helicity suppressed. 
Apart from adding a whole new type of interaction to the process, NP might also arise from new 
axial vector interactions, such that 
$C_{10} = \mathcal{C}_{10}^{\mathrm{SM}} + \mathcal{C}_{10}^{\mathrm{NP}}$, which is currently the 
main sensitivity test related to the latest anomalies observed \cite{rk}. Hence, a small 
difference between the measured and the theoretical values of the branching fraction 
might be a strong hint of NP \cite{DeBruyn,Fleischer}.

\section{Experimental beauty}
An improved measurement of the $B_{(s)}^0\rightarrow\mu^+\mu^-$ branching fractions has been 
performed by LHCb using the full Run 1 and Run 2 data \cite{PRL,PRD}. To perform the analysis, 
a pair of opposite charged muons with invariant mass, 
$m_{\mu^+\mu^-} \in [4900, 6000] \mathrm{ MeV/c}^2$, are selected by requiring them 
to form a good-quality vertex displaced from the interaction point. 

\begin{figure}[h]
       \centerline{\includegraphics[scale=0.18]{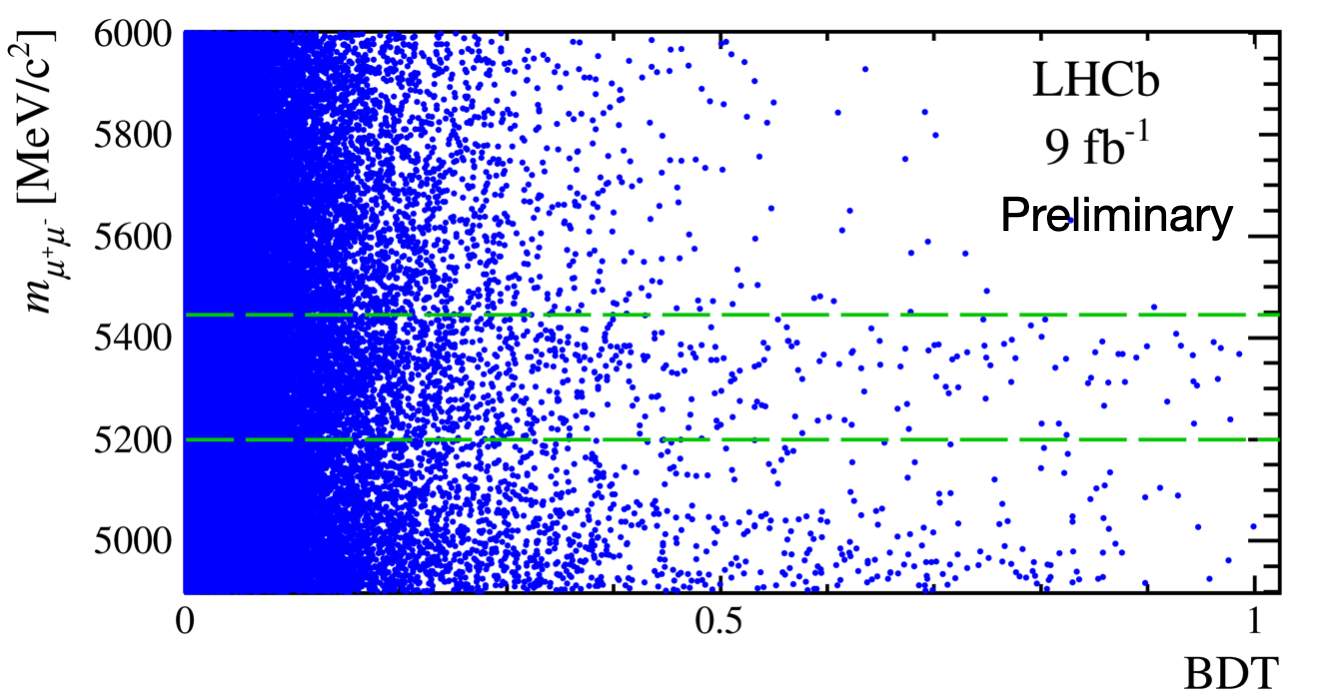}}
       \caption{Distribution of events for the full dataset in BDT and invariant mass of the muons. 
       The dashed green lines represent the signal mass region.}
       \label{fig:BDT_scatter}
\end{figure}

The signal yields are obtained from a maximum likelihood fit to the dimuon invariant mass. 
To increase the separation between signal and background, the fit is performed in 
bins of a Boosted Decision Tree (BDT) classifer, which categorizes events as more or less 
signal-like. This is done by assigning an score between 0 and 1 to each event such that, 
the closer the score to 0, the more background-like the event is. The BDT distribution 
of events for the full Run 1 and Run 2 datasets is shown in Fig.~\ref{fig:BDT_scatter} 
where the dash green lines represent the limits of the signal mass region.  

From the signal yields the branching fractions are determined using 
$B^+ \rightarrow J/\psi K^+$ and $B^0\rightarrow K^+\pi^-$ as normalization channels, 
the former having similar particle identification (PID) and trigger requirements 
as the signal and the latter, similar kinematics.

Due to the limited statistics on the signal sample, several calibrations and estimations 
have to be performed to constrain the fit parameters, such as the estimation of the 
shapes and yields of the various backgrounds components, the signal shape calibration 
and the BDT calibration among others. The yields of the exclusive backgrounds are 
estimated from a combination of theoretical inputs, simulation and techniques based on data. 

In addition, the mean and the width of the $B_{(s)}^0\rightarrow\mu^+\mu^-$ mass shapes, both 
described by a Double-Sided Crystal Ball (DSCB) function \cite{DSCB}, are also calibrated. 
For the mean, fits to the invariant mass of the $B_s^0 \rightarrow K^+ K^-$ and 
$B^0 \rightarrow K^+ \pi^-$ data samples are performed using a DSCB as the mass shape. 
Fits to the charmonium ($J/\psi$, $\psi(2S)$) and bottonium 
($\Upsilon(1S)$, $\Upsilon(2S)$, $\Upsilon(3S)$) resonances decaying to two muons are used to 
calibrate the width of the signal shape, which is determined from the interpolation 
$\sigma_{\mu\mu} (m_{\mu\mu}) =a_0 + a_1\cdot m_{\mu\mu}$, where the $a_0$ and $a_1$ are 
the interpolation constants. Finally, the tail parameters of the DSCB signal 
shapes are obtained from simulation samples, which are smeared with the 
resolution determined from the previously mentioned $\bar{q}q$ resonances.

\subsection{BDT calibration}
The $B_{(s)}^0\rightarrow\mu^+\mu^-$ decays 
are extremely rare processes and the measurements of their branching fractions are statistically 
limited. Therefore, the sensitivity of the data sample should be increased as much as possible. 
This is achieved by dividing the data sample in subsets of the BDT score and performing the 
mass fit simulatenously in all these subsets. Several binning schemes have been studied 
among which the best performant one has been found to be a six BDT bins scheme with 
the limits 0, 0.25, 0.4, 0.5, 0.6, 0.7 and 1.

\begin{figure}[h]
       \centerline{\includegraphics[scale=0.19]{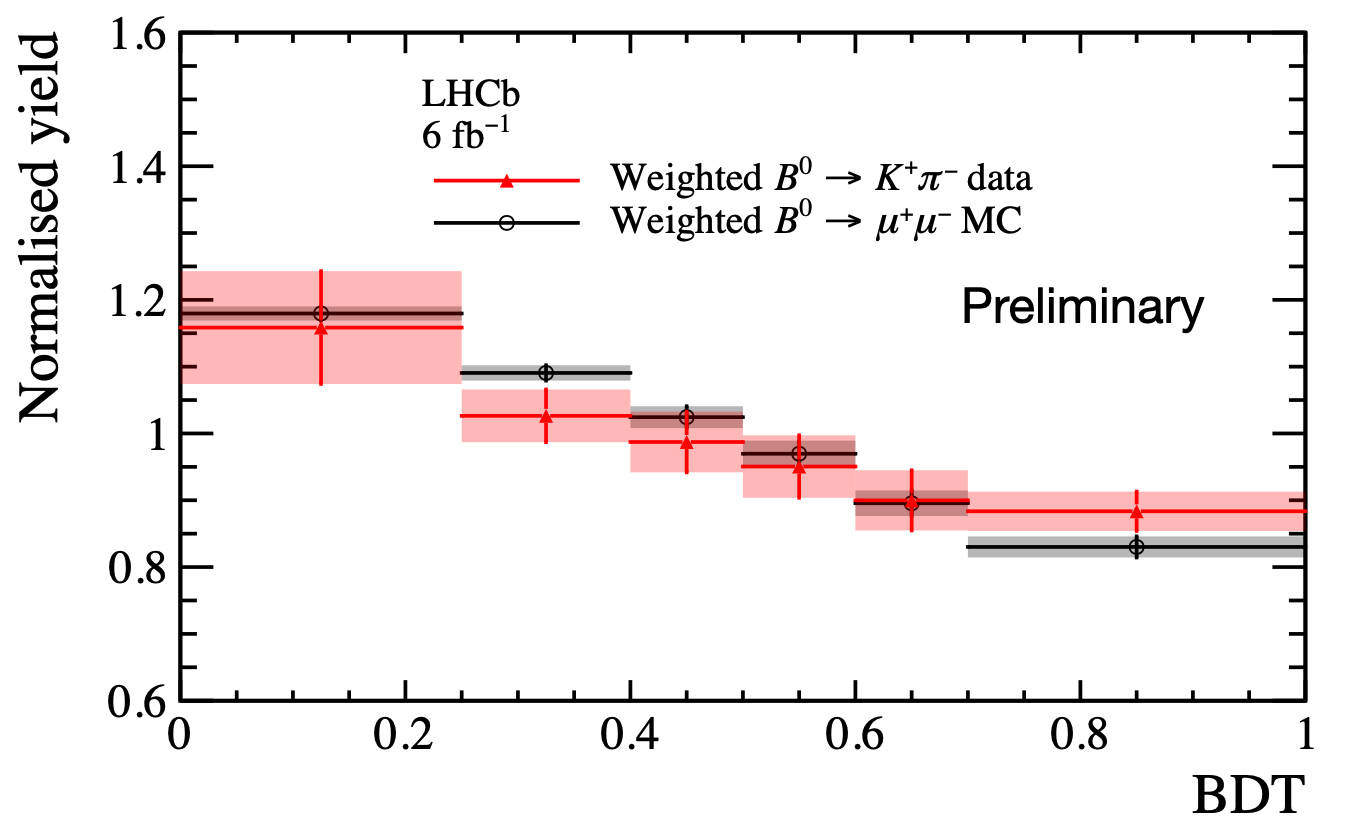}}
       \caption{Expected BDT distribution for $B^0\rightarrow\mu^{+} \mu^{-}$ candidates using the 
       two methods: in black $B^0 \rightarrow \mu^{+} \mu^{-}$ weighted simulation and 
       in red $B^0\rightarrow K^+\pi^-$ corrected data.}
       \label{fig:BDT_calibration}
\end{figure}

As the fit is performed in bins of the BDT, the expected relative yield of 
$B_{(s)}^0\rightarrow\mu^+\mu^-$ events per BDT bin should be estimated, 
a procedure known as BDT calibration. The BDT calibration can be performed on corrected data, 
using the normalization channel $B^0\rightarrow K^+\pi^-$ corrected for the trigger 
and PID requirements, or on weighted simulation, directly from a 
$B^0_{(s)}\rightarrow\mu^+\mu^-$ simulation sample weighted to take into account the 
simulation and data differences. Fig.~\ref{fig:BDT_calibration} displays the estimated 
yields of $B^0\rightarrow\mu^+\mu^-$ for Run 2 obtained following the two different 
approaches. While the former strategy was followed in the previous analysis \cite{lhcbprev}, 
this time the latter has been analysed and used. As shown in Fig.~\ref{fig:BDT_calibration}, 
both strategies are in very good agreement and the uncertainty on the relative yields determined 
from the weighted simulation is significantly smaller with respect to results obtained from the 
previous strategy. This uncertainty reduction on the BDT calibration has a direct impact on the 
final uncertainty of the branching fraction measurement. 

\subsection{Final fit and results}
\begin{figure}
       \begin{minipage}{0.5\linewidth}
       \centerline{\includegraphics[scale=0.18]{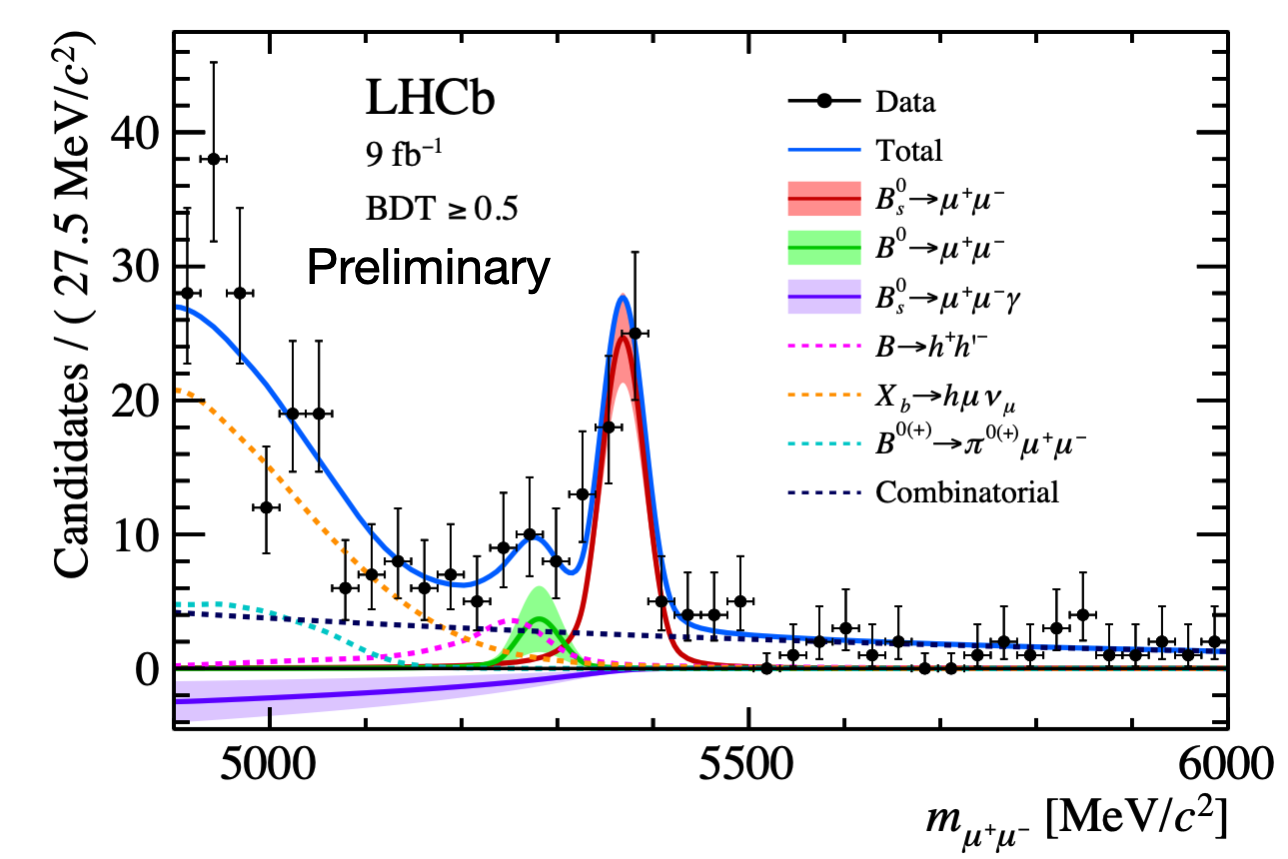}}
       \caption{Mass distribution of $B^0_{s}\rightarrow\mu^{+} \mu^{-}$, 
       $B^0\rightarrow\mu^{+} \mu^{-}$ candidates with BDT $>$ 0.5 in red and green, 
       respectively. The result of the fit is shown in blue and the separated components 
       for the combinatorial and exclusive backgrounds are also detailed.}
       \label{fig:fit}
       \end{minipage}
       \hfill
       \begin{minipage}{0.4\linewidth}
       \centerline{\includegraphics[scale=0.18]{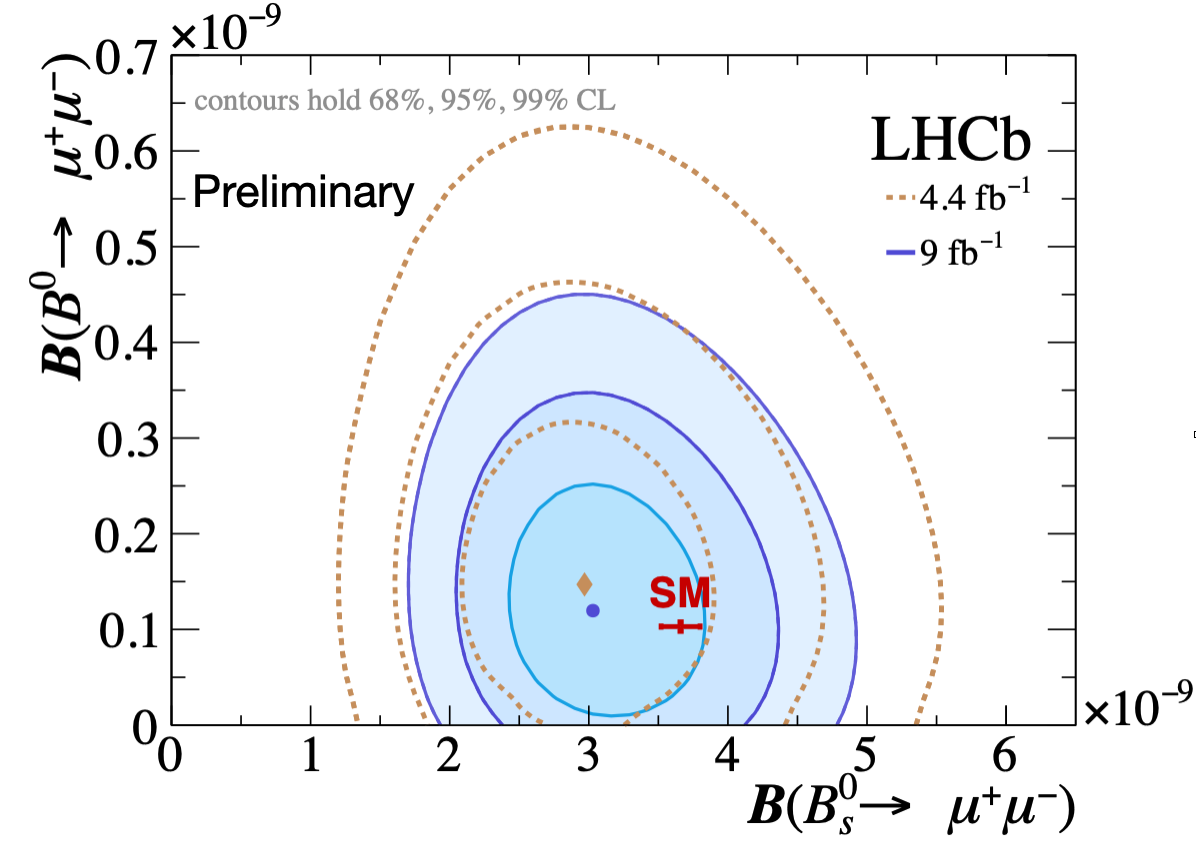}}
       \caption{Two dimensional representation of the measured braching fraction
        values of $B^0\rightarrow\mu^{+} \mu^{-}$ and $B^0_s\rightarrow\mu^{+} \mu^{-}$. 
        The contours of the previous analysis are represented in yellow and the SM 
        value is displayed in red}
       \label{fig:contours}
       \end{minipage}
\end{figure}

The BDT calibration together with more ingredients, such as the signal calibration or the background 
estimation briefly mentioned before, are taken into account in the final invariant mass fit by 
constraining some of its parameters and shapes. Fig.~\ref{fig:fit} displays 
the mass distribution of the $B^0\rightarrow\mu^{+} \mu^{-}$ and $B^0_{s}\rightarrow\mu^{+} \mu^{-}$ 
candidates with BDT larger than 0.5 in green and red, respectively. In the same figure, the 
exclusive and combinatorial background distributions are also shown. From the fit, the branching 
fraction of $B^0_{s}\rightarrow\mu^{+} \mu^{-}$ is determined while a stringent upper limit is 
set for the $B^0\rightarrow\mu^{+} \mu^{-}$ branching fraction:
\bea
\mathcal{B}(B_s^0\to\mu^+\mu^-) &= &(3.09^{+ 0.46+0.15}_{-0.43-0.11}) \times 10^{-9} \nonumber\\
\mathcal{B}(B^0\to\mu^+\mu^-) &< & 2.6 \times 10^{-10} \mbox{ at 95\% CL}\nonumber
\eea
where the first and second uncertainties on the $B_s^0\to\mu^+\mu^-$ branching fraction correspond 
to the statistical and systematic uncertainties, respectively. The values measured are in good 
agreement with the previous results and with the SM predictions. This can also be seen from 
the two dimensional representation of both measured branching fractions illustrated in 
Fig.~\ref{fig:contours}, where the contour from the previous analysis and the SM value 
are represented in yellow and red, respectively. 

An important difference with respect to the previous analysis is the addition of the search of 
$B_s^0\to\mu^+\mu^-\gamma$, whose fit component is shown in purple in Fig.~\ref{fig:fit}, 
a process for which an upper limit on its branching fraction of 
$\mathcal{B}(B_s^0\to\mu^+\mu^-\gamma)_{m_{\mu\mu}>4.9\mbox{ }\mathsf{ GeV}/c^2} < 2.0 \times 10^{-9}$ 
at 95\% CL has been established. 

\section{Conclusion}
The latest results for the branching fractions of $B_{(s)}^0\to\mu^+\mu^-$ decays have been 
presented. The branching fraction of $B_{s}^0\to\mu^+\mu^-$ is determined to be 
$\mathcal{B}(B_s^0\to\mu^+\mu^-) = (3.09^{+ 0.46+0.15}_{-0.43-0.11}) \times 10^{-9}$, where 
the first and second uncertainties correspond to the statistical and systematic uncertainties, 
respectively. This is the most precise single experiment measurement of the rarest of the 
B-hadron processes measured so far. An upper limit of 
$\mathcal{B}(B^0\to\mu^+\mu^-) <  2.6 \times 10^{-10} \mbox{ at 95\% CL}$ is set 
for the $B^0\to\mu^+\mu^-$ branching fraction, where no significant excess is found. Similarly, 
an upper limit is found for the $B_s^0\to\mu^+\mu^-\gamma$ process of 
$\mathcal{B}(B_s^0\to\mu^+\mu^-\gamma)_{m_{\mu\mu}>4.9\mbox{ }\mathsf{ GeV}/c^2} < 2.0 \times 10^{-9}$. 
All results are consistent with the SM predictions and previous results \cite{lhcbprev,combination} 
and help to further constraint NP models.

\end{document}